\documentclass[journal,twoside,web]{ieeecolor}
\usepackage{jsen}
\usepackage{cite}
\usepackage{amsmath,amssymb,amsfonts}
\usepackage{algorithmic}
\usepackage{graphicx}
\usepackage{textcomp}
\usepackage{wrapfig}
\def\BibTeX{{\rm B\kern-.05em{\sc i\kern-.025em b}\kern-.08em
    T\kern-.1667em\lower.7ex\hbox{E}\kern-.125emX}}
\usepackage{graphicx}
\usepackage{epstopdf}
\usepackage[process=auto]{pstool} 
\usepackage{psfrag}

\usepackage{hyperref}

\usepackage{bm}

\begin{document}
\title{FootSLAM meets Adaptive Thresholding}
\author{Johan Wahlstr\"{o}m, Andrew Markham, and Niki Trigoni
	\thanks{This research has been financially supported by the National Institute of Standards and Technology (NIST) via the grant \emph{Pervasive, Accurate, and Reliable Location-based Services for Emergency Responders} (Federal Grant: 70NANB17H185).}
\thanks{J. Wahlstr\"om, A.Markham and N. Trigoni are with the Department of Computer Science, University of Oxford, Oxford, UK (\{johan.wahlstrom, andrew.markham, niki.trigoni\}@cs.ox.ac.uk).}}

\IEEEtitleabstractindextext{\begin{wrapfigure}[12]{r}{3in}%
\vspace*{-5mm}
\hspace*{2mm}
\includegraphics[width=2.6in]{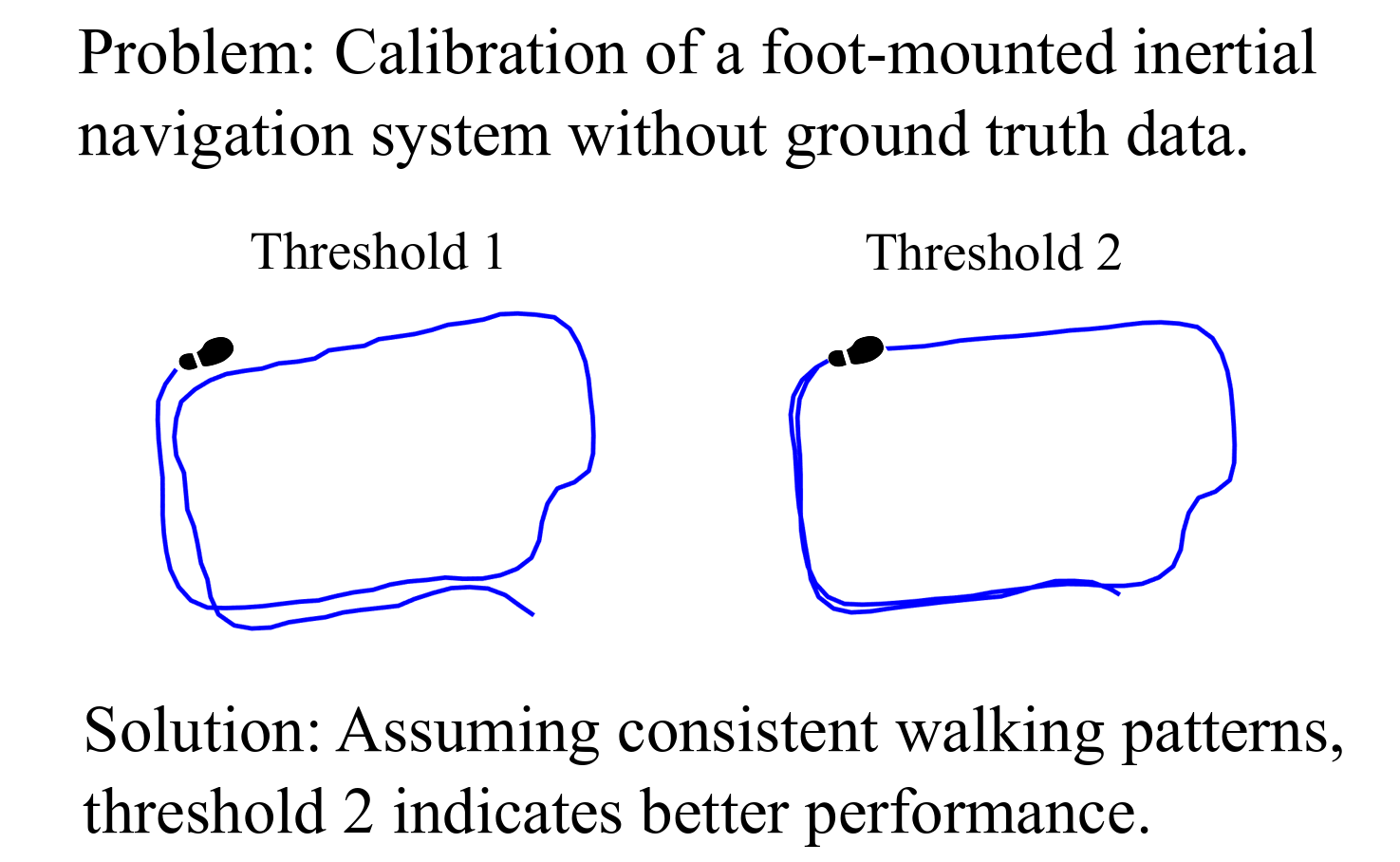}%
\end{wrapfigure}%
\begin{abstract}
Calibration of the zero-velocity detection threshold is an essential prerequisite for zero-velocity-aided inertial navigation. However, the literature is lacking a self-contained calibration method, suitable for large-scale use in unprepared environments without map information or pre-deployed infrastructure. In this paper, the calibration of the zero-velocity detection threshold is formulated as a maximum likelihood problem. The likelihood function is approximated using estimation quantities readily available from the FootSLAM algorithm. Thus, we obtain a method for adaptive thresholding that does not require map information, measurements from supplementary sensors, or user input. Experimental evaluations are conducted using data with different gait speeds, sensor placements, and walking trajectories. The proposed calibration method is shown to outperform fixed-threshold zero-velocity detectors and a benchmark using a speed-based threshold classifier. 
\end{abstract}

}

\maketitle

\section{Introduction}

Inertial navigation aided by zero-velocity updates (ZUPT) has been hailed as one of the most promising technologies for indoor positioning in environments without pre-installed infrastructure or prior map information. Consider, for example, firefighters arriving at an emergency scene with low visibility, intense heat, scattered debris and building materials, and no general knowledge of the area. In this situation, ZUPT-aided inertial navigation provides a reliable, low-cost positioning solution with no setup time and no dependence on environmental conditions such as visibility or obstacles in line-of-sight \cite{Zhang2013,Wahlstrom2019b}. Other relevant applications include gaming \cite{Zhou2016}, biomedicine \cite{Mccarthy2019}, military positioning \cite{Rantakokko2011}, and analysis of sports performance \cite{Falbriard2018}.

The performance of a ZUPT-aided inertial navigation system (INS) is highly dependent on the design and calibration of the zero-velocity detector. Typically, the detector is implemented as a generalized likelihood ratio test; the sensor unit is considered to be stationary if the likelihood ratio exceeds a user-specified threshold \cite{Skog2010}. If the threshold on the likelihood ratio is too large, the detector will not be able to detect stationary instances when the user is running. If the threshold is too small, the detector will produce false zero-velocity instances. In addition, the optimal threshold will be dependent on factors such as gait technique, the placement of the sensor, the type of shoe, and the walking surface. 

\begin{figure}[t]
	\def\svgwidth{4.25in}
	\vspace*{0.5mm}
	\hspace*{-2.5mm}
	\scalebox{0.81}{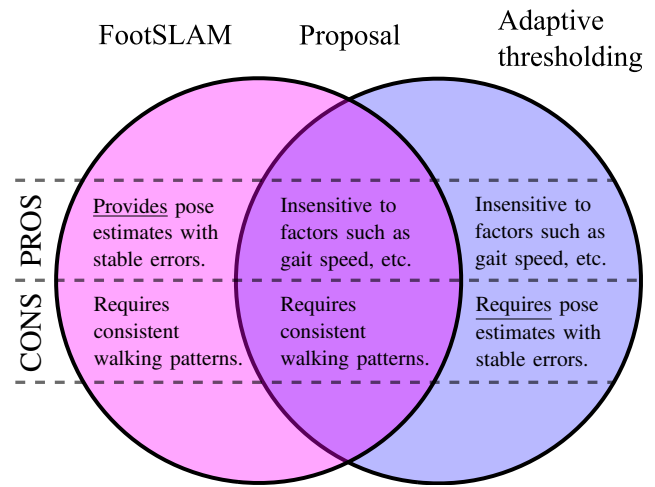}
	\caption{\label{venn_figure}A Venn diagram illustrating the relation between FootSLAM, previous methods for adaptive thresholding, and our proposal.}
	\vspace*{-2.5mm}
\end{figure}

Adaptive thresholding has been explored in several studies \cite{Tian2016,Park2016,Zhang2017,Ma2017,Wagstaff2017,Wagstaff2019,Wahlstrom2019,Wang2019,Wang2015}. Often, the threshold is set based on the result of a speed or motion mode classification. However, the predefined threshold values, associated with the respective motion modes, need to be calibrated using ground truth position data. Since such data is not available in unprepared environments, this means that current implementations of adaptive thresholding require an extensive calibration period -- separate from the real-world deployment for which the navigation system is intended -- and that the threshold values, once set, cannot adapt to real-time changes in gait or environment conditions that were not accounted for in the calibration process.

As illustrated in Fig. \ref{venn_figure}, the problem of calibrating the zero-velocity detection threshold is in this paper approached via the FootSLAM algorithm, a type of simultaneous localization and mapping (SLAM) algorithm. FootSLAM is a method for transforming odometry with position drift into pose estimates with long-term error stability. Thus, the calibration can be performed by treating the output from FootSLAM as a pseudo ground truth. The calibration algorithm is formulated as the solution to a maximum likelihood problem, and the likelihood function is approximated using particle weights produced by FootSLAM. In this way, a joint calibration and navigation algorithm that is completely independent of any supplementary ground truth data is obtained. The method is validated using a diverse set of experimental data. 


Sections \ref{section_inertial_odometry} and \ref{section_footslam} review previous work on adaptive thresholding and FootSLAM, respectively. Section \ref{section_calibration_using_footslam} describes the proposed algorithm. Experimental results are presented in Section \ref{section_experimental_results} and the article is concluded in Section \ref{section_summary}.

\section{Inertial Odometry and Adaptive Thresholding}
\label{section_inertial_odometry}

The purpose of a ZUPT-aided INS is to compute the odometry estimates
\begin{equation}
\mathbf{z}_{1:T}={\cal G}_{\boldsymbol{\theta}}(\mathbf{y}_{1:T}).
\end{equation}
Here, $\mathbf{z}_k$ denotes an estimate of the three-dimensional translation and rotation of the inertial sensors between sampling instances $k-1$ and $k$. Further, $\mathbf{y}_k$ denotes the inertial measurements at sampling instance $k$, $\mathbf{y}_{1:T}\overset{_\Delta}{=}\{\mathbf{y}_1,\dots,\mathbf{y}_T\}$, and the transformation ${\cal G}_{\boldsymbol{\theta}}(\cdot)$ is a filter or smoother composed of the navigation equations and a zero-velocity detector with the zero-velocity detection threshold $\boldsymbol{\theta}$. The transformation ${\cal G}_{\boldsymbol{\theta}}(\cdot)$ is dependent on several design parameters, including initialization parameters, parameters characterizing sensor errors, and parameters used by the zero-velocity detector. However, we will, in similarity with previous studies on parameter estimation for ZUPT-aided INSs, primarily focus on the tuning of the zero-velocity detection threshold \cite{Tian2016,Park2016,Zhang2017,Ma2017,Wagstaff2017,Wagstaff2019,Wahlstrom2019}. The optimal threshold value may vary with a large number of factors, and an improper tuning can have a detrimental effect on performance.

To find a suitable threshold, one must typically make use of ground truth data in the form of maps, user provided location information, or measurements from complementary sensors. A common approach is to first estimate or classify the speed or motion mode of the user. Based on the result, the detector selects a threshold value that has been optimized, using ground truth data, for that specific speed or motion class \cite{Tian2016,Park2016,Zhang2017,Ma2017,Wagstaff2017,Wagstaff2019}. However, other calibration methods have also been proposed. In \cite{Wahlstrom2019}, a time-varying threshold was obtained by formulating the likelihood ratio test in a Bayesian setting; in \cite{Liu2014}, the threshold was set based on the variance of the accelerometer measurements computed over a specified time window; and in \cite{Wang2018}, the threshold was fixed while instead varying the window length of the samples used to compute the detection statistic. There have also been attempts at designing robust zero-velocity detectors by using neural networks \cite{Wagstaff2018}, \cite{Yu2019}, by incorporating velocity estimates into the detector \cite{Walder2010,Ren2016}, or by inferring the state of gait cycle \cite{Li2012,Sun2018}.

\section{Footslam} 
\label{section_footslam}

\begin{figure}[t]
	\hspace*{-1.5mm}
	\vspace*{2mm}
	{
		\includegraphics{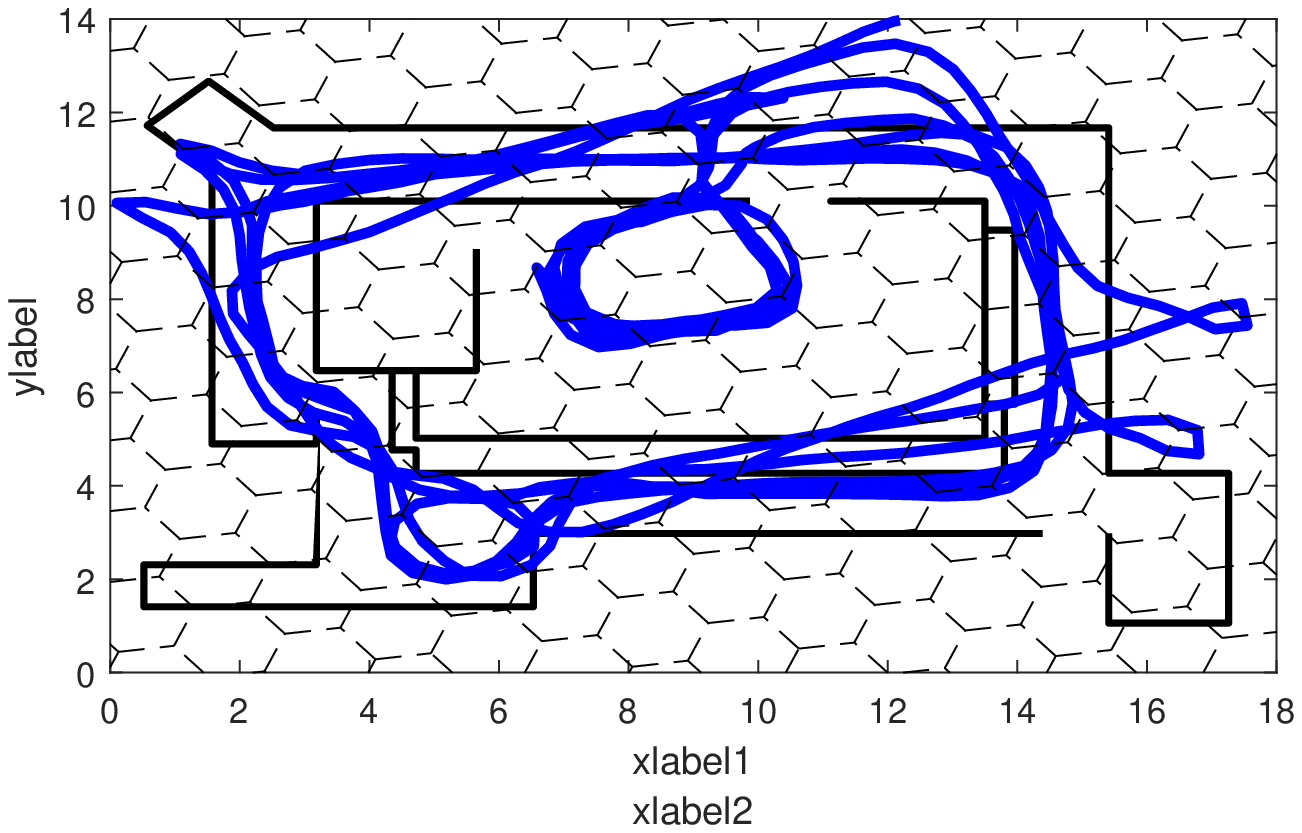}
		\vspace*{0mm}
	}
	\hspace*{-2.2mm}
	{
		\vspace*{-1.3mm}
		\includegraphics{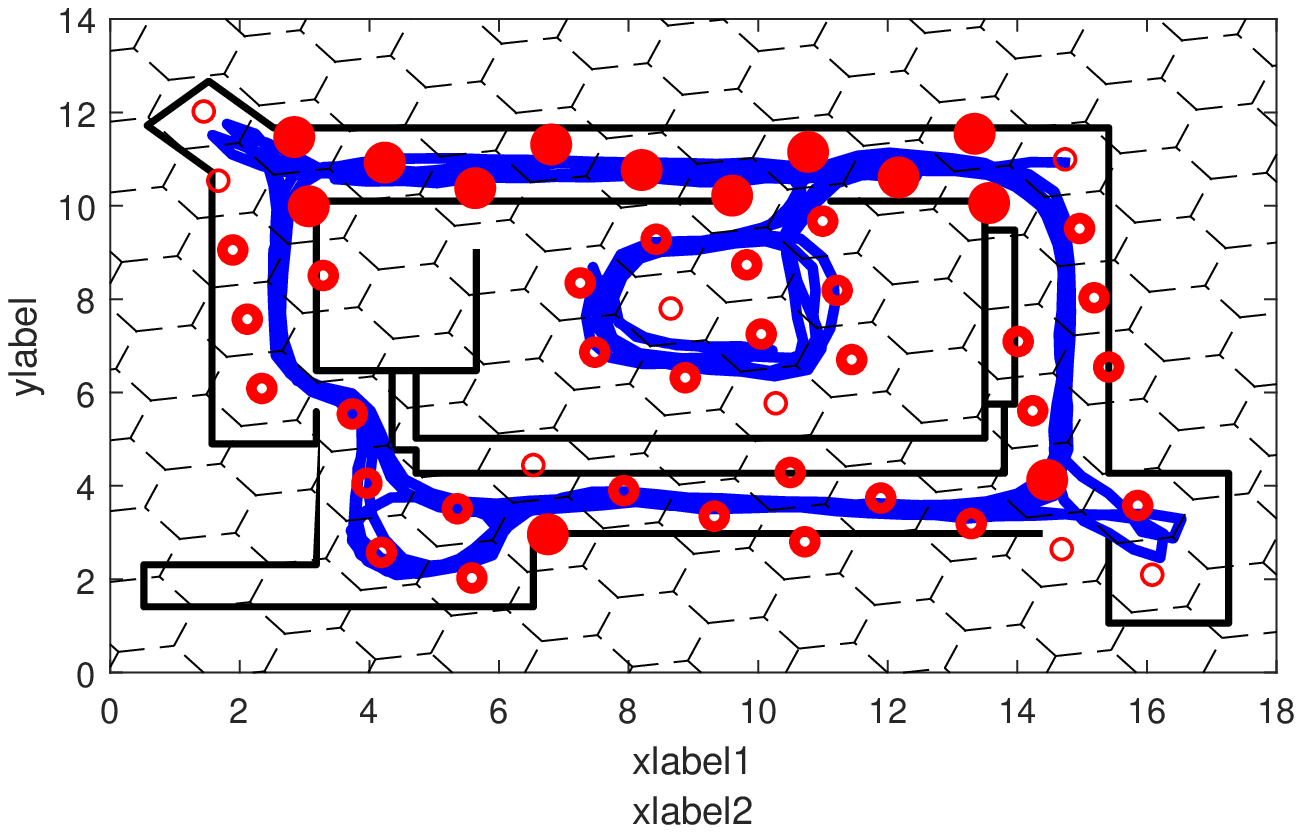}
		\vspace*{0mm}
	}
	\vspace*{-3mm}
	\caption{Illustration of (a) stand-alone zero-velocity-aided inertial navigation and (b) FootSLAM after about four minutes of walking in an office building. The sizes of the red circles in (b) indicate the prevalence of transitions involving the associated hexagons.}
	\vspace*{0mm}
	\label{fig_footslam_illustration}
\end{figure}

The idea of FootSLAM is to represent a two-dimensional navigation area using a grid of hexagons, and then learn the probability of transitioning from a given hexagon to an adjacent one \cite{Robertson2009}. This is illustrated in Fig. \ref{fig_footslam_illustration}. The inference framework utilizes the Rao-Blackwellized particle filtering approach of the FastSLAM algorithm. Thus, the posterior  $p(\mathbf{x}_{0:T},\mathbf{m}|\mathbf{z}_{1:T})$ is factorized as
\begin{equation}
p(\mathbf{x}_{0:T},\mathbf{m}|\mathbf{z}_{1:T})=\underbrace{p(\mathbf{m}|\mathbf{x}_{0:T})}_{\text{map estimation}}\cdot \underbrace{p(\mathbf{x}_{0:T}|\mathbf{z}_{1:T})}_{\text{pose estimation}}
\end{equation}
where $\mathbf{x}$ and $\mathbf{m}$ represent the pose and map, respectively, with $\mathbf{z}_k$ treated as a noisy measurement of the difference between $\mathbf{x}_{k-1}$ and $\mathbf{x}_k$. The pose is recursively estimated according to
\begin{align}
p(\mathbf{x}_{0:k}|\mathbf{z}_{1:k})\propto& \underbrace{p(\mathbf{z}_k|\mathbf{x}_{k-1:k})}_{\text{ likelihood}}\hspace*{0mm}\underbrace{p(\mathbf{x}_k|\mathbf{x}_{0:k-1})}_{\text{pose transition}}\hspace*{0mm}\underbrace{p(\mathbf{x}_{0:k-1}|\mathbf{z}_{1:k-1})}_{\text{previous posterior}}.
\end{align} 
The likelihood function $p(\mathbf{z}_k|\mathbf{x}_{k-1:k})$ is used to draw samples of $\mathbf{x}_k$, whereas the pose transition probability $p(\mathbf{x}_k|\mathbf{x}_{0:k-1})$, which is computed by marginalizing over the map, is used in the particle weight update
\begin{eqnarray}
\label{eq_particle_weight_update}
w_k^{(i)}\propto p(\mathbf{x}_k^{(i)}|\mathbf{x}_{0:k-1}^{(i)}) w_{k-1}^{(i)}
\end{eqnarray}
where $w^{(i)}$ denotes the weight of the $i$th particle.
The pose transition probability is large when $\mathbf{x}_{k-1}\rightarrow\mathbf{x}_{k}$ or $\mathbf{x}_{k}\rightarrow\mathbf{x}_{k-1}$ corresponds to a frequently observed hexagon transition, and the filter will thus favor particles that revisit the same hexagon transitions (and thereby outline consistent walking patterns). However, note that without any absolute heading, position, or scale information, the estimates are invariant under rotation, translation, and scaling of the odometry in the plane.

Finally, the navigation solution is represented by a set of pose estimates $\{\{\mathbf{x}_k^{(i)}\}_{i=1}^N\}_{k=0}^T$ and associated weights $\{\{w_k^{(i)}\}_{i=1}^N\}_{k=0}^T$,
where $N$ is the number of particles. Extensions of the FootSLAM algorithm have considered estimation of systematic odometry errors \cite{Robertson2009}, navigation and mapping in three dimensions \cite{Puyol2014}, collaborative mapping \cite{Puyol2011}, fusion with magnetic field measurements \cite{Robertson2013} and user provided hints \cite{Angermann2012}, and navigation in the presence of moving platforms such as escalators and elevators \cite{Kaiser2018}. 

\begin{figure*}[t]
	\def\svgwidth{1.93\columnwidth}
	\hspace*{4mm}
	\scalebox{1}{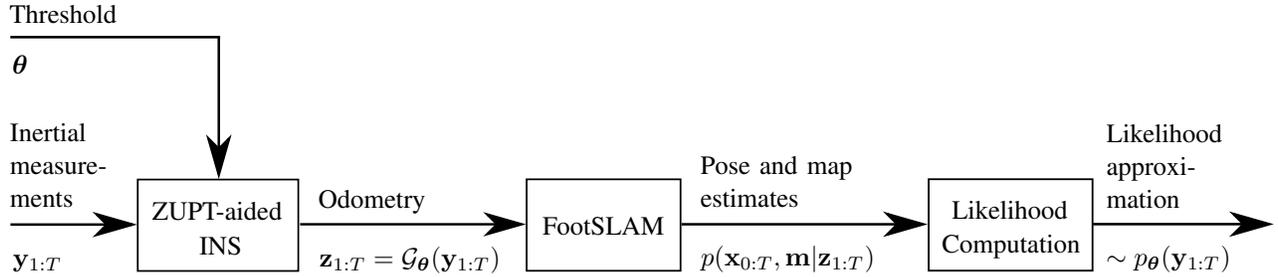}
	\vspace*{0mm}
	\caption{System overview. The FootSLAM algorithm is fed with odometry from a ZUPT-aided INS. The likelihood is approximated using estimates provided by FootSLAM.}
	\label{fig_schematic_illustration}
	\vspace*{-0mm}
\end{figure*}

\section{Adaptive Thresholding using Footslam}
\label{section_calibration_using_footslam}

The maximum likelihood estimate of the threshold is
\begin{equation}
\boldsymbol{\hat{\theta}}=\underset{\boldsymbol{\theta}}{\arg\max}\;p_{\boldsymbol{\theta}}(\mathbf{y}_{1:T})
\end{equation}
where the likelihood function can be approximated as
\begin{align}
\begin{split}
\label{eq_likelihood_approximation}
p_{\boldsymbol{\theta}}(\mathbf{y}_{1:T})&\approx p(\mathbf{z}_{1:T})\\
&=\textstyle\prod_{k=1}^T
p(\mathbf{z}_{k}|\mathbf{z}_{1:k-1})\\
&=\textstyle\prod_{k=1}^T\int p(\mathbf{z}_k|\mathbf{x}_{k-1:k})p(\mathbf{x}_k|\mathbf{x}_{0:k-1})\\
&\hspace*{13mm}\cdot p(\mathbf{x}_{0:k-1}|\mathbf{z}_{1:k-1})d\mathbf{x}_{0:k}\\
&\approx\textstyle\prod_{k=1}^T\sum_{i=1}^Np(\mathbf{x}_k^{(i)}|\mathbf{x}_{0:k-1}^{(i)})w_{k-1}^{(i)}
\end{split}
\end{align}
and we use the convention that $\mathbf{z}_{1:0}=\emptyset$. The first approximation in \eqref{eq_likelihood_approximation} corresponds to	approximations made in the nonlinear system for ZUPT-aided inertial navigation and when only using the point estimates of the odometry, whereas the second approximation is a conventional particle filter approximation \cite{Olsson2008}. 

The result in \eqref{eq_likelihood_approximation} demonstrates that the value of the likelihood function for a given threshold can be approximated based on the output obtained from FootSLAM when using the same threshold to compute the odometry. In particular, comparing with \eqref{eq_particle_weight_update}, we see that the likelihood approximation in \eqref{eq_likelihood_approximation} uses the sum of the particle weights \textit{before} normalization. Thus, as should be intuitive, the likelihood function is large when $p(\mathbf{x}_k^{(i)}|\mathbf{x}_{0:k-1}^{(i)})$ tends to be large, i.e., when the particles are prone to make repeated hexagon transitions. 

By utilizing the recursion 
\begin{align}
\label{eq_likelihood_recursion}
p_{\boldsymbol{\theta}}(\mathbf{y}_{1:k})\approx p_{\boldsymbol{\theta}}(\mathbf{y}_{1:k-1})\cdot\textstyle\sum_{i=1}^Np(\mathbf{x}_k^{(i)}|\mathbf{x}_{0:k-1}^{(i)})w_{k-1}^{(i)}
\end{align}
with the initialization $p_{\boldsymbol{\theta}}(\mathbf{y}_{1:0})=1$, the value of the likelihood function can be updated after each time step in the FootSLAM algorithm, and there's no need to store all pose and map estimates produced by FootSLAM. Further, note that the likelihood value can be computed online by updating the ZUPT-aided INS, the FootSLAM algorithm, and the likelihood estimates after obtaining each new sample of inertial measurements. The relation between ZUPT-aided inertial navigation, FootSLAM, and the likelihood computation is illustrated in Fig. \ref{fig_schematic_illustration}. In the end, we find an approximate maximum likelihood estimate by performing a grid search over a specified set of threshold values\footnote{The set of thresholds can in practice be quite limited in size. An upper bound for the threshold can be obtained by requiring the detector to produce a detected zero-velocity instance with a high probability when the inertial sensors are perfectly stationary. Likewise, the threshold cannot be arbitrarily small, since this would mean that the foot would be very far off from being stationary when ZUPTs are applied. Given the empirical relationship between the odometry performance and the threshold (see Figs. \ref{fig_rectangle_speed} (a) and \ref{fig_placement} (a)), we have found it sufficient to consider about $M=25$ thresholds with a logarithmic spacing on the logarithm of the threshold.}. The method for estimating the threshold is summarized in Algorithm 1. 

The computational complexity of the proposed algorithm is comparable to that of the FootSLAM algorithm (the computational resources required by the ZUPT-aided INS are negligible in comparison). The article \cite{Puyol2013a} studied the
computational complexity of the FootSLAM algorithm in detail and demonstrated that FootSLAM can be used for real-time applications. 

\begin{table}[t]
	\normalsize
	\rule{251pt}{0.5pt}\\[-10pt]
	\rule{251pt}{0.5pt}
	\textbf{Algorithm 1:} Maximum likelihood estimation of the zero-velocity detection threshold. \\[-6pt]
	\rule{251pt}{0.5pt}
	\textit{Input}: Inertial measurements $\mathbf{y}_{1:T}$. \\
	\textit{Output}: Threshold estimate $\boldsymbol{\hat{\theta}}$.
	\\[-6pt]
	\rule{251pt}{0.5pt}
	\begin{enumerate}
		\item Specify a set of thresholds $\{\boldsymbol{\theta}^{(1)},\dots,\boldsymbol{\theta}^{(M)}\}$. For $j=1,\dots,M$:
		\begin{itemize}
			\item[a)] Compute the odometry $\mathbf{z}_{1:T}$
			by applying a ZUPT-aided INS with the threshold $\boldsymbol{\theta}^{(j)}$ to the inertial measurements $\mathbf{y}_{1:T}$.
			\item[b)] Run FootSLAM on the odometry $\mathbf{z}_{1:T}$ and approximate $p_{\boldsymbol{\theta}^{(j)}}(\mathbf{y}_{1:T})$ using \eqref{eq_likelihood_recursion}.
		\end{itemize}
		\item Choose the estimate as the threshold that produced the largest value of the likelihood function. \\ [-18pt]
	\end{enumerate}
	\rule{251pt}{0.5pt}\\[-10pt]
	\rule{251pt}{0.5pt}
\end{table}

%

\section{Experimental Results}
\label{section_experimental_results}

The proposed method was evaluated in two separate experiments. The experiments demonstrate calibration during both repetitive walking along a marked trajectory and during day-to-day walking in an office environment. In addition, we considered several different gait speeds and sensor placements. Inertial measurements were collected at a sampling rate of $100\,[\text{Hz}]$ from a Xsens MTi-3-8A7G6-DK IMU. The odometry was computed using a Kalman smoother \cite{Colomar2012}, implemented with the stance hypothesis optimal detection (SHOE) zero-velocity-detector \cite{Skog2010}. The yaw rate bias was included as a state element in the FootSLAM algorithm \cite{Angermann2012}, and the particles were resampled using systematic resampling \cite{Hol2006}.
All data recordings started and ended at the same position. Therefore, the difference between the initial and final position estimates was used to evaluate the position error of the final position estimate. Since the position errors of a ZUPT-aided INS primarily stem from yaw errors (and not scaling errors) \cite{Nilsson2012}, this is a suitable and convenient method for evaluation\footnote{Refer to \cite{Nilsson2013} for related details on the observability of a ZUPT-aided INS.}. \emph{The data and the code used in the experiments are available at \href{https://www.cs.ox.ac.uk/people/johan.wahlstrom/}{\textcolor{blue}{https://www.cs.ox.ac.uk/people/johan.wahlstrom/}}}.

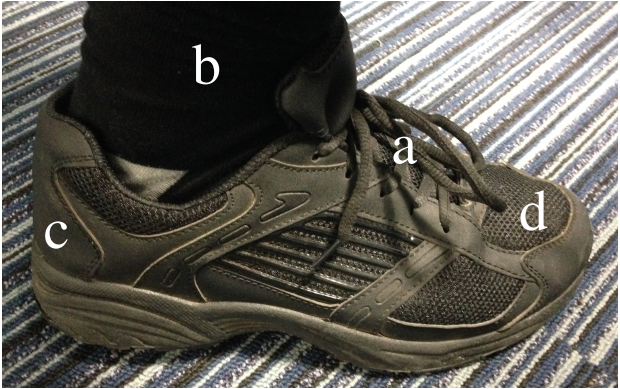
\begin{figure}[t]
	\def\svgwidth{0.42\columnwidth}
	\hspace*{2mm}
	\scalebox{2.2}{
\begingroup%
  \makeatletter%
  \providecommand\color[2][]{%
    \errmessage{(Inkscape) Color is used for the text in Inkscape, but the package 'color.sty' is not loaded}%
    \renewcommand\color[2][]{}%
  }%
  \providecommand\transparent[1]{%
    \errmessage{(Inkscape) Transparency is used (non-zero) for the text in Inkscape, but the package 'transparent.sty' is not loaded}%
    \renewcommand\transparent[1]{}%
  }%
  \providecommand\rotatebox[2]{#2}%
  \newcommand*\fsize{\dimexpr\f@size pt\relax}%
  \newcommand*\lineheight[1]{\fontsize{\fsize}{#1\fsize}\selectfont}%
  \ifx\svgwidth\undefined%
    \setlength{\unitlength}{212.5984252bp}%
    \ifx\svgscale\undefined%
      \relax%
    \else%
      \setlength{\unitlength}{\unitlength * \real{\svgscale}}%
    \fi%
  \else%
    \setlength{\unitlength}{\svgwidth}%
  \fi%
  \global\let\svgwidth\undefined%
  \global\let\svgscale\undefined%
  \makeatother%
  \begin{picture}(1,0.62666667)%
    \lineheight{1}%
    \setlength\tabcolsep{0pt}%
    \put(0,0){\includegraphics[width=\unitlength,page=1]{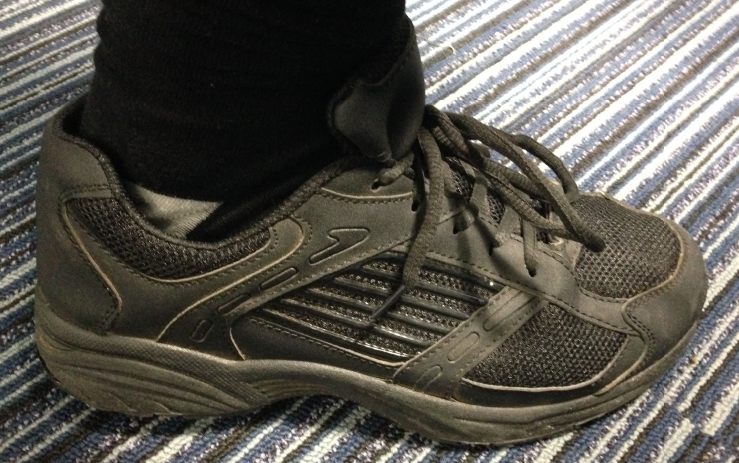}}%
    \put(0.63194855,0.36381714){\color[rgb]{0,0,0}\makebox(0,0)[lt]{\lineheight{1.25}\smash{\begin{tabular}[t]{l}\textcolor{white}{a}\end{tabular}}}}%
    \put(0.30836947,0.49311646){\color[rgb]{0,0,0}\makebox(0,0)[lt]{\lineheight{1.25}\smash{\begin{tabular}[t]{l}\textcolor{white}{b}\end{tabular}}}}%
    \put(0.0673969,0.22607015){\color[rgb]{0,0,0}\makebox(0,0)[lt]{\lineheight{1.25}\smash{\begin{tabular}[t]{l}\textcolor{white}{c}\end{tabular}}}}%
    \put(0.83940887,0.25555705){\color[rgb]{0,0,0}\makebox(0,0)[lt]{\lineheight{1.25}\smash{\begin{tabular}[t]{l}\textcolor{white}{d}\end{tabular}}}}%
  \end{picture}%
\endgroup%
}
	\vspace*{0mm}
	\caption{Illustration of the sensor placements a) `Shoelaces', b) `Ankle', c) `Heel', and d) `Toes'.}
	\label{fig_illustration_of_placements}
	\vspace*{-0mm}
\end{figure}

\subsection{Closed-loop Trajectory with Speed Variations}
\label{section_closed_loop}

\begin{figure}[t]
	\hspace*{-1mm}
	\vspace*{0mm}
	{
		\includegraphics{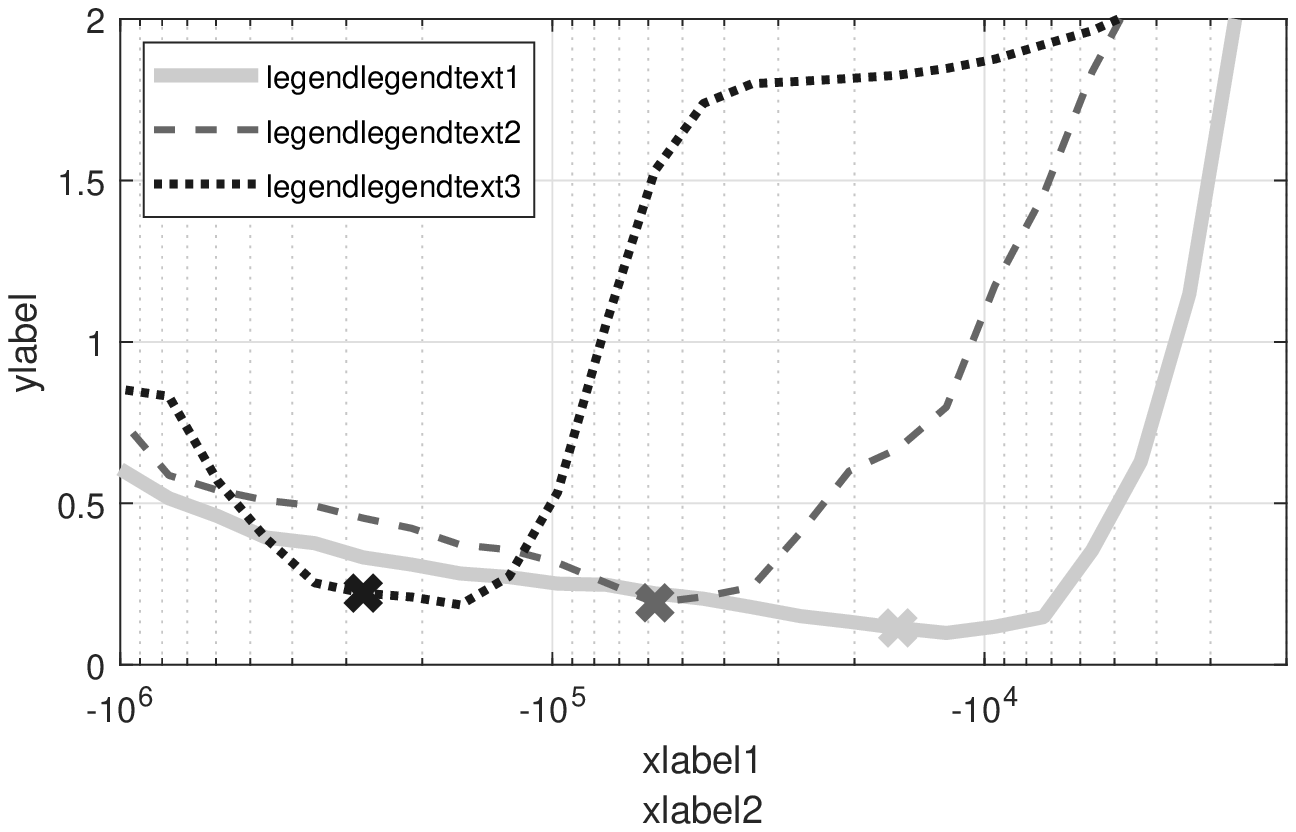}
		\vspace*{-2mm}
	}

	{	\hspace*{-0mm}
		\vspace*{0mm}
		\includegraphics{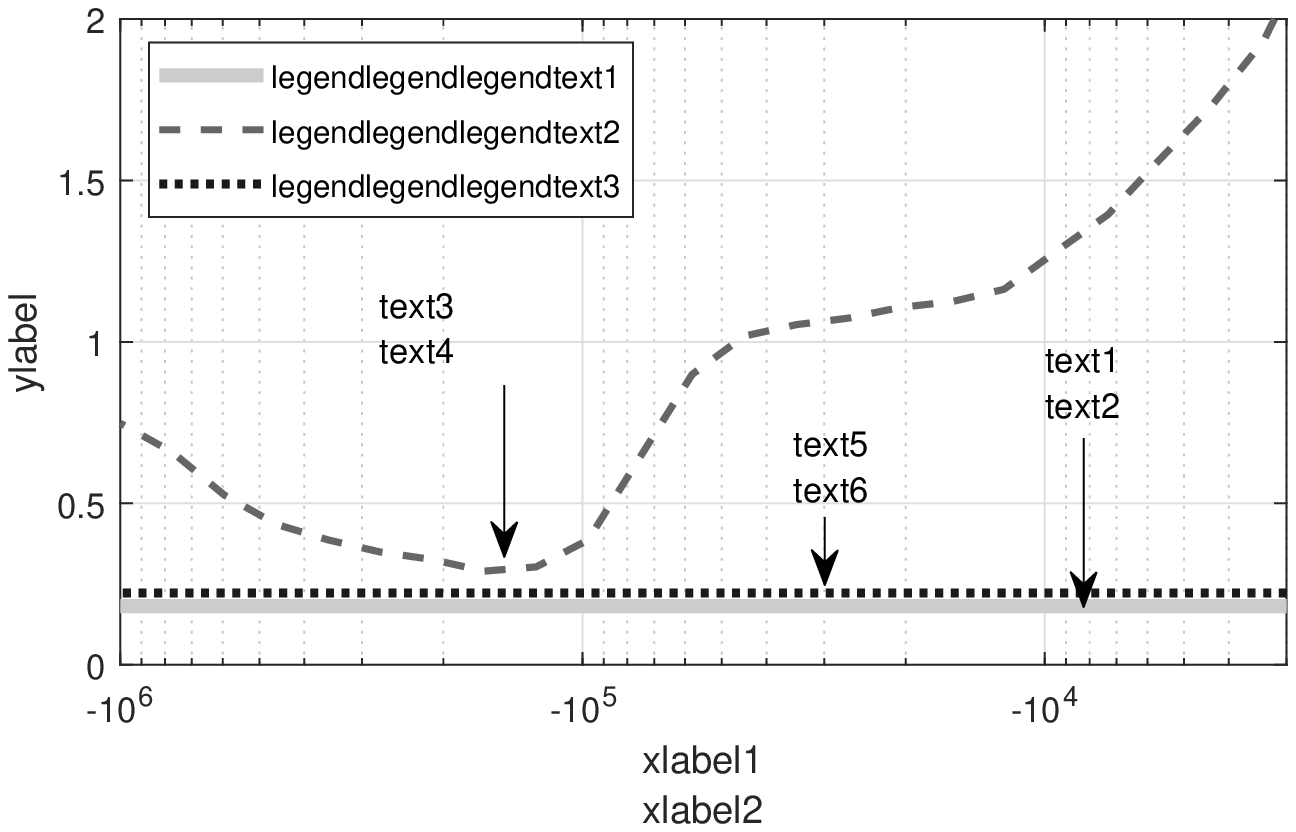}
	}
	\vspace*{-6mm}
	\caption{Position error of a ZUPT-aided INS after walking along a closed-loop trajectory with a length of about twelve meters. The crosses in a) indicate the thresholds chosen by the calibration algorithm. Three gait modes were used: walking, fast walking, and jogging.}
	\vspace*{0mm}
	\label{fig_rectangle_speed}
\end{figure}

In the first experiment, calibration and evaluation data were collected for three gait modes: walking, fast walking, and jogging, with average speeds of about $4.5\,[\text{km/h}]$, $6.5\,[\text{km/h}]$, and $8\,[\text{km/h}]$, respectively. The sensor was placed on top of the shoelaces as illustrated by placement a) in Fig. \ref{fig_illustration_of_placements}. All data was recorded while walking or jogging along a rectangle of dimensions $2.6\,[\text{m}]\times3.2\,[\text{m}]$. For each gait mode, we collected i) calibration data consisting of one data recording of ten consecutive laps, and ii) evaluation data consisting of 50 data recordings of one lap each.  Fig. \ref{fig_rectangle_speed} (a) displays the position root-mean-square error (RMSE) for the three gait modes, computed using the initial and final position estimates in each recording in the evaluation data.

\subsubsection{Benchmarks}

The most common approach to adaptive thresholding is to choose the threshold based on the result of a speed or motion mode classification. 
Many variants of this idea have been explored \cite{Tian2016,Park2016,Zhang2017,Ma2017,Wagstaff2017,Wagstaff2019}. In the first experiment, the proposed method was benchmarked against a speed-based classifier. The latter was implemented as follows. For each gait mode, the FootSLAM algorithm was applied to the calibration data. The position estimates produced by FootSLAM were then used as pseudo ground truth when evaluating the accuracy of the odometry for different thresholds. The optimal threshold for each speed was chosen as the threshold with the minimal time-averaged position RMSE. Following this, each data recording in the evaluation data was classified based on a user-calibrated rule-based classifier. Specifically, the classification was performed based on the average speed while in movement. All recordings with an average speed exceeding $7.5\,[\text{km/h}]$ were classified as jogging, and all recordings with an average speed below $5.5\,[\text{km/h}]$ were classified as walking. The threshold was then chosen accordingly among the three optimized threshold values. Given the characteristics of the calibration data (i.e., walking or jogging along a clearly defined rectangle), the FootSLAM algorithm will perform reliably, and the performance of the speed-based classifier will not suffer from inaccuracies in the pseudo ground truth computed using FootSLAM. Rather, the performance of the speed-based threshold will primarily be limited by the quantity of calibration data and the extent to which the calibration data reflects the evaluation data. 

Note that the proposed algorithm does not use any speed or motion mode classification. Rather, it makes the assumption that all conditions affecting the optimal choice of threshold are stationary long enough for FootSLAM to converge and to make use of the results. Thus, when evaluating the proposed algorithm, all evaluation data associated with a given gait mode employed the threshold optimized using the calibration data from the same gait mode. A discussion on how to extend the proposal to include speed or motion mode classification is included in Section \ref{section_summary}. In addition to the speed-based classifier, the proposed method was also compared against a fixed-threshold detector. All three algorithms explored the same discrete set of $M=25$ thresholds.

\begin{figure}[t]
	\vspace*{0mm}
\hspace*{-0.2mm}
\includegraphics{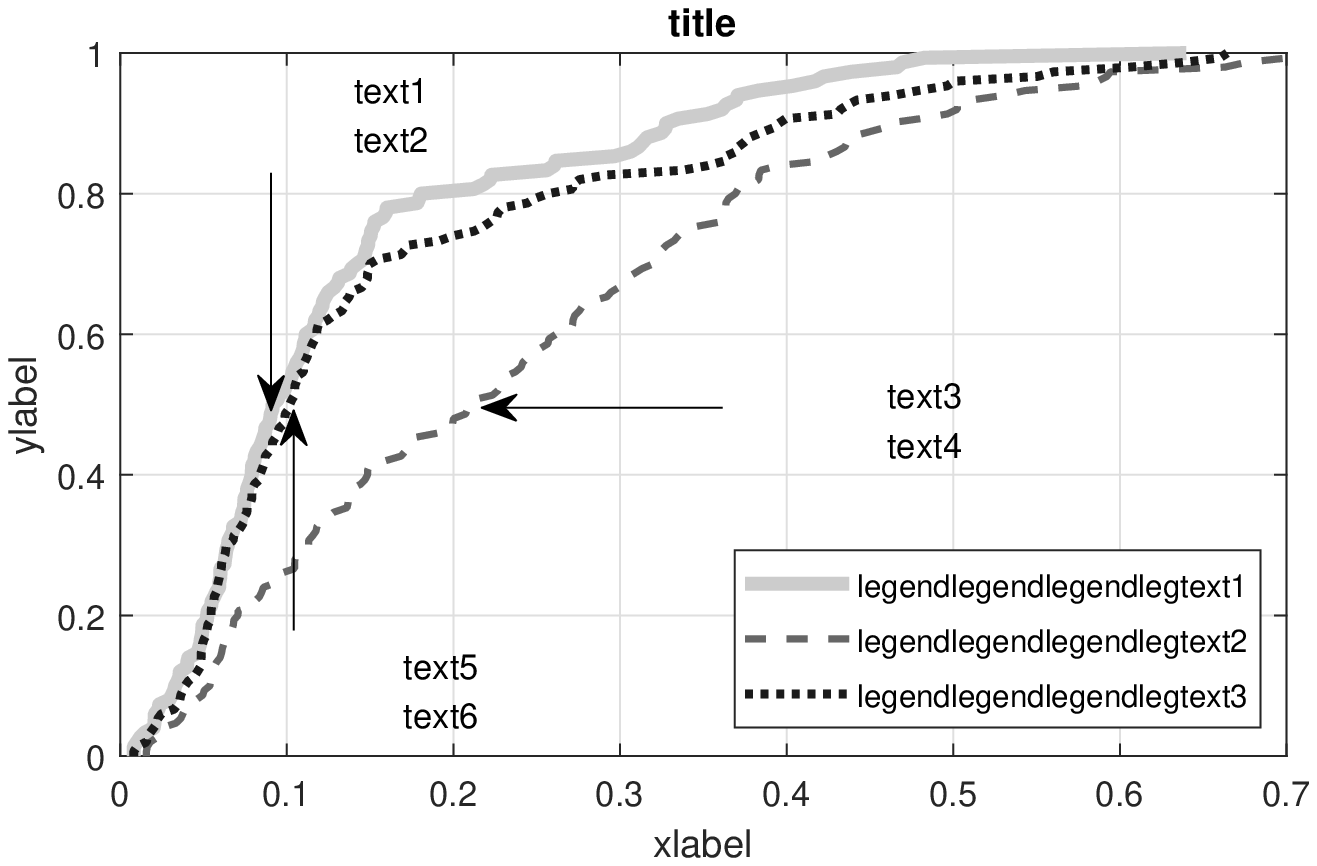}
\vspace*{-6mm}
	\caption{Empirical cumulative distribution functions of position errors after walking along a closed-loop trajectory with a length of about twelve meters. Three gait modes were used: walking, fast walking, and jogging.}
	\vspace*{0mm}
	\label{fig_ecdf}
\end{figure}

\subsubsection{Results}

Optimal thresholds were estimated using the calibration data. The evaluation data was then used to compare the performance of ZUPT-aided INSs using 
\begin{itemize}
	\item for each gait mode, the threshold estimate found with the method described in Section \ref{section_calibration_using_footslam} (denoted ``adaptive threshold'' in the figures).
	\item for each gait mode, the threshold estimate found using the speed-based classifier described in the present section (denoted ``benchmark'').
	\item the same threshold for all gait modes (denoted ``fixed threshold'').
\end{itemize}
The resulting horizontal position RMSE is shown in Fig. \ref{fig_rectangle_speed} (b)\footnote{Note that the horizontal axis displays the value of the fixed threshold. Thus, the RMSEs for the adaptive threshold and the benchmark, which are not dependent on the fixed threshold, are shown as a horizontal lines.}. As can be seen, the adaptive threshold performs significantly better than the best fixed threshold and somewhat better than the benchmark. Fig. \ref{fig_ecdf} compares the adaptive threshold, the benchmark, and the best (in terms of RMSE) fixed threshold, by displaying the empirical cumulative distribution functions (ECDF) of the horizontal position error. In comparison with the best fixed threshold, the adaptive threshold reduces the median horizontal position error by more than 50$\%$. When concatenating all evaluation data into a single trajectory of length $1.74\,[\text{km}]$, the norm of the horizontal position error of the adaptive threshold, the benchmark, and best fixed threshold becomes $16.58\,[\text{m}]$, $19.68\,[\text{m}]$, and $31.48\,[\text{m}]$, respectively.

\subsubsection{Comments on the Performance Comparison} 

When examining the results, it is important to take into consideration that the performance of methods that rely on ground truth data, such as the benchmark, will be highly dependent on the quantity and quality of the training data. Therefore, it is hard to make a fair quantitative comparison between the such methods and the proposed algorithm, which does not require ground truth data. However, the important qualitative differences can be  summarized as follows. On the one hand, the proposed method benefits from not requiring ground truth training data and being able to calibrate against variations in any factor that will influence the threshold calibration. On the other hand, the proposed method is limited by its requirement on convergence of the FootSLAM algorithm during the calibration phase.

\subsection{Office Environment with Different Sensor Placements}

\begin{figure}[t]
	\hspace*{-0.2mm}
	\includegraphics{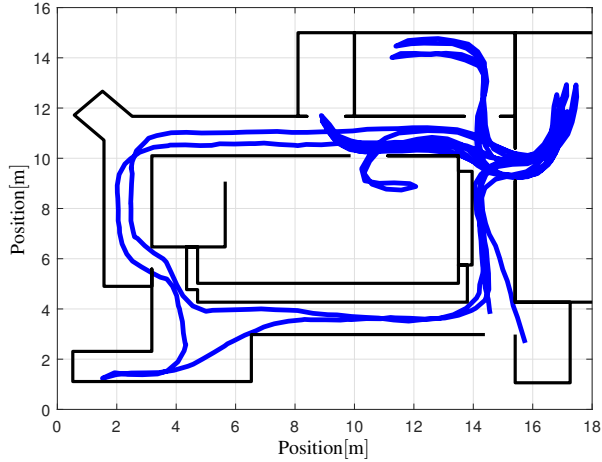}
	\vspace*{-7mm}
	\caption{Trajectory used for calibration in office environment.}
	\label{fig_office_illustration}
\end{figure}

\begin{figure}[t]
	\vspace*{0mm}
	\hspace*{-1mm}
	{
		\includegraphics{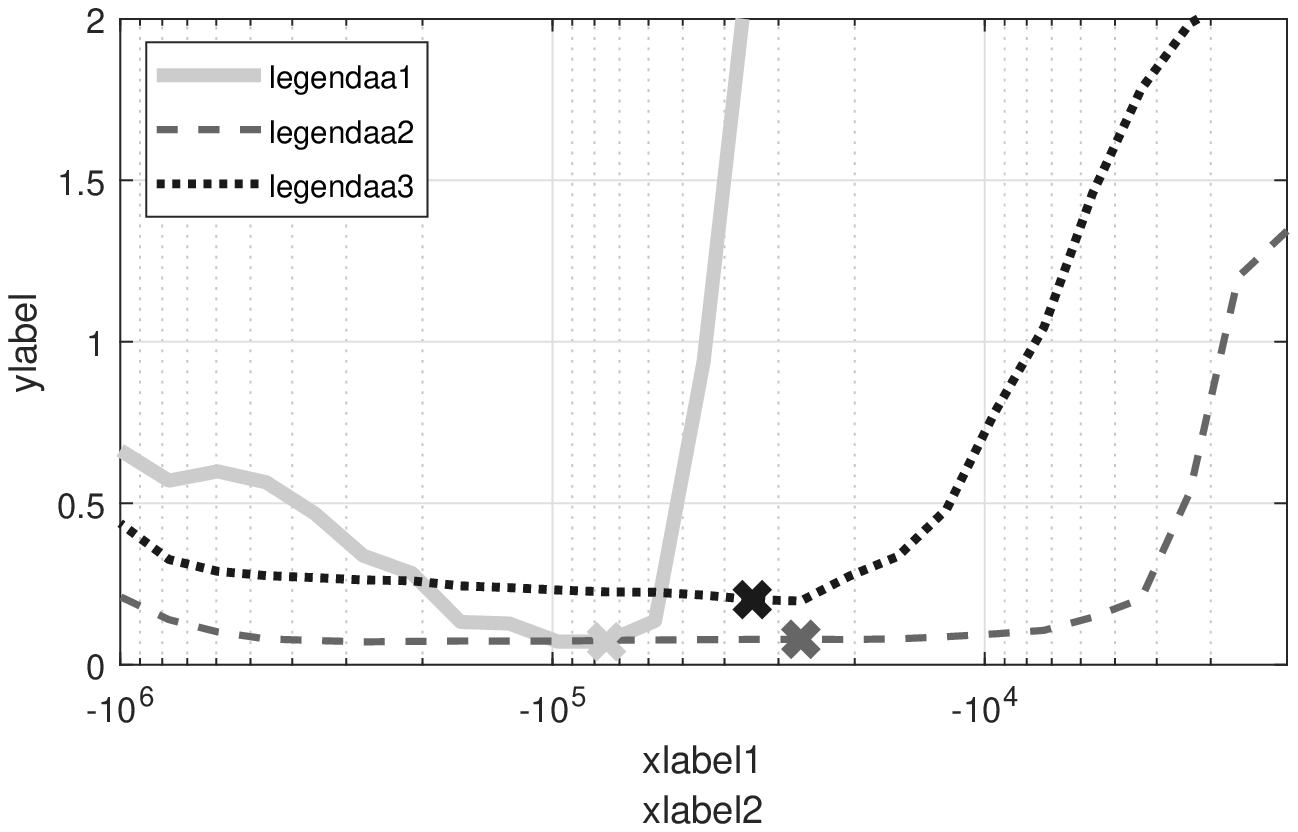}
		\vspace*{-2mm}
	}

	{	\hspace*{0mm}
		\vspace*{0mm}
		\includegraphics{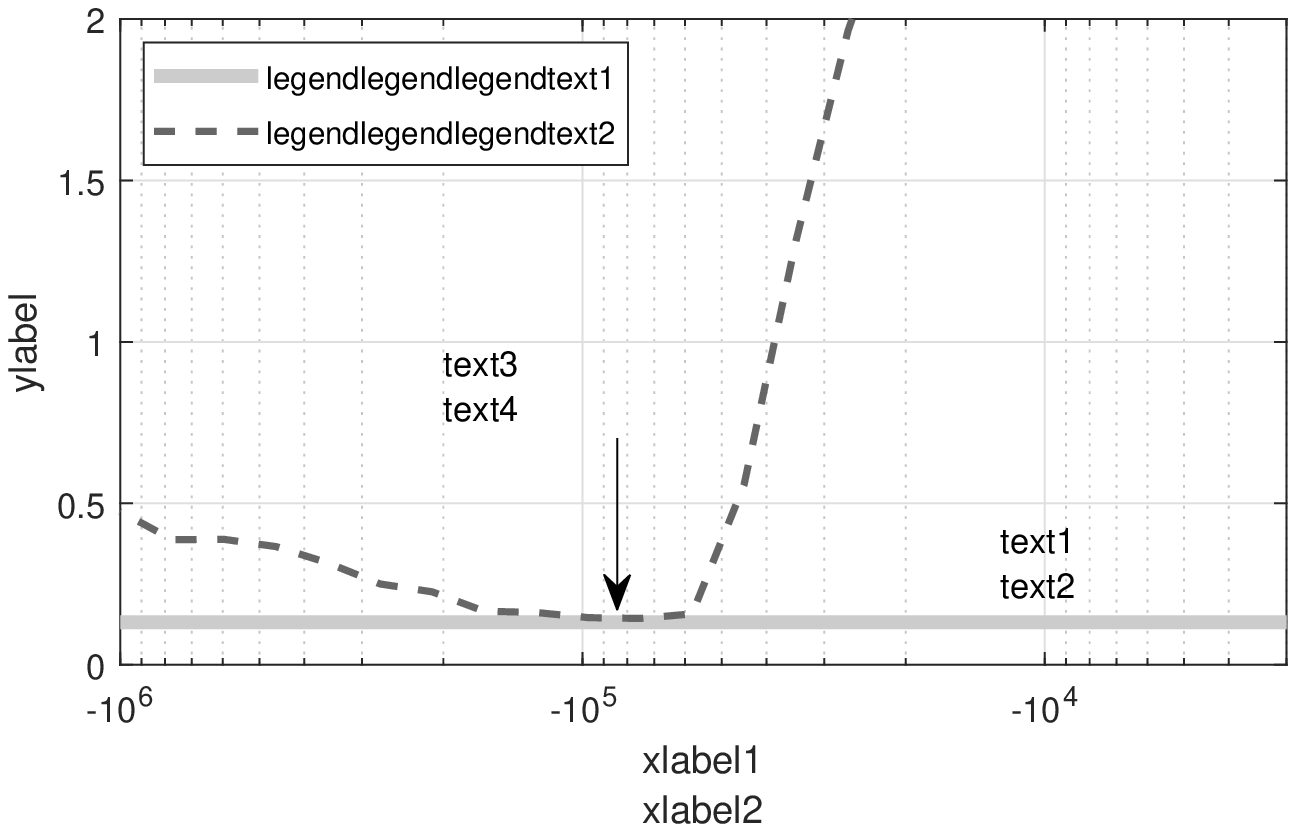}
	}
	\vspace*{-6mm}
	\caption{Position error of a ZUPT-aided INS after walking along a closed-loop trajectory with a length of about twelve meters. The crosses in a) indicate the thresholds chosen by the proposed calibration algorithm. The data was recorded using the three sensor placements `Ankle', `Heel', and `Toes', illustrated in Fig. \ref{fig_illustration_of_placements}.}
	\vspace*{0mm}
	\label{fig_placement}
\end{figure}

\begin{figure}[t]
	\vspace*{0mm}
	\hspace*{-0.2mm}
	\includegraphics{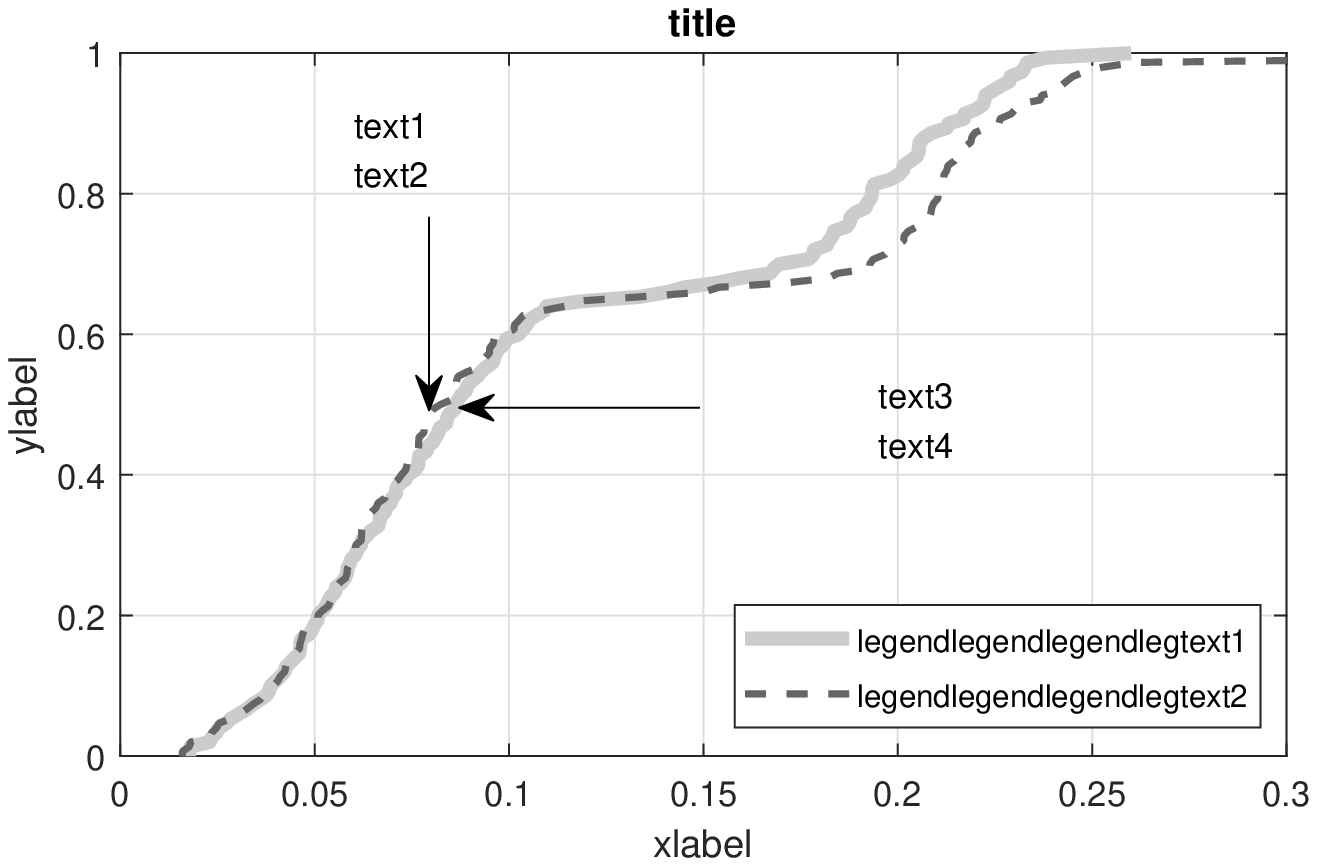}
	\vspace*{-6mm}
	\caption{Empirical cumulative distribution functions of position errors after walking along a closed-loop trajectory with a length of about twelve meters. The data was recorded using the three sensor placements `Ankle', `Heel', and `Toes',
	illustrated in Fig. \ref{fig_illustration_of_placements}.}
	\vspace*{0mm}
	\label{fig_ecdf_placement}
\end{figure}

\textcolor{black}{In the second experiment, calibration and evaluation data were recorded using the three sensor placements `Ankle', `Heel', and `Toes', illustrated in Fig. \ref{fig_illustration_of_placements}. The gait speed was around $6\,[\text{km/h}]$. The calibration data was recorded while walking in an office environment of about $200\,[\text{m}^2]$. The pedestrian started at the entrance door and then walked for three minutes in between his personal desk, a meeting room, a kitchen, a lab room, a printer, and a bathroom. The trajectory is illustrated in Fig. \ref{fig_office_illustration}. The evaluation data was recorded in the same way as in the first experiment. Fig. \ref{fig_placement} (a) displays the resulting position RMSE for the three sensor placements\footnote{The speed-based benchmark considered in the first experiment was not used here since the data did not contain any significant speed variations}. As seen from Figs. \ref{fig_placement} (b) and \ref{fig_ecdf_placement}, the adaptive threshold gives a slight performance improvement in comparison with the best fixed threshold. However, when interpreting these results, note that the two sensor placements `Heel' and `Toes' perform well for a wide range of thresholds. As a result, even an optimal choice of threshold (in terms of RMSE) for each sensor placement would only improve the RMSE by about one and a half centimeter in comparison with the best fixed threshold. When merging all evaluation data, the norm of the horizontal position error of the adaptive threshold and best fixed threshold becomes $13.16\,[\text{m}]$ and $14.03\,[\text{m}]$, respectively. The position errors obtained when concatenating all evaluation into a single trajectory are summarized in Table \ref{Tab1}.}


\begin{table}[t]
	\normalsize
	\caption{norm of position error after merging all evaluation data. \label{Tab1}}
	\centering
	\begin{tabular}{l|cc}
		\hline
		\hline
		&  \multicolumn{2}{l}{} \\ [-2.3ex]
		& \multicolumn{2}{c}{Threshold} \\
		\cline{2-3} 
		&  \multicolumn{2}{l}{} \\ [-2.3ex]
		Varying factor & Adaptive & Best fixed\\
		&  \multicolumn{2}{l}{} \\ [-2.7ex]
		\hline    
		&  \multicolumn{2}{l}{} \\ [-2.3ex]
		Gait speed & $16.58\,[\text{m}]$ & $31.48\,[\text{m}]$ \\
		Sensor placement & $13.16\,[\text{m}]$ & $14.03\,[\text{m}]$ \\
		\hline \hline 
	\end{tabular}
	\vspace*{-0mm}
\end{table}

%

\section{Summary and Discussion}
\label{section_summary}

\color{black}

This paper has demonstrated the complementary benefits of two research directions -- FootSLAM and adaptive thresholding -- that for a long period of time has developed separately. There are two problems with established methods for adaptive thresholding. Firstly, they need to be trained using large sets of data. Secondly, they can typically only adjust to variations in speed or gait mode, and not to variations in for example the walking surface or the sensor placement. However, by calibrating a ZUPT-aided INS using the position estimates provided by FootSLAM, it is possible to solve both of these problems. Specifically, we have presented a maximum likelihood-based algorithm for adaptive thresholding that is completely independent of ground truth data or additional information. A short description of the algorithm follows. The likelihood function is evaluated for a given set of thresholds. For each threshold, the corresponding odometry is computed, which is then used to run the FootSLAM algorithm. Finally, the likelihood function is approximated using the output from the FootSLAM algorithm. Using data with varying gait speeds and sensor placements, the resulting adaptive detector was shown to outperform detectors with fixed thresholds as well as a benchmark detector that classified the evaluation data based on the estimated speed. Under speed variations, the median horizontal position error was reduced by more than 50$\%$ in comparison with a fixed-threshold detector. Note that the presented algorithm is different from a calibration based only on loop closure error since 1) the algorithm does not require ground truth data or other supplementary information for loop-closure detection; and 2) the calibration is based on repeated hexagon transitions in FootSLAM rather than on loop closures per se.

The only price to pay for not relying on ground truth data is that the FootSLAM algorithm needs to converge during the calibration phase. In other words, there needs to be a period of time with a) a reasonably fixed threshold-performance relationship; and b) consistent (in terms of hexagon transitions) walking patterns. However, as supported by the large number of publications on the FootSLAM algorithm, physical constraints provided by walls or other obstacles are in many situations sufficient to enforce this consistency in the walking patterns. In addition, when intentionally walking according to consistent motion patterns for a limited period of time, the proposed algorithm enables calibration in e.g., outdoor areas where it may be difficult to set up infrastructure-dependent sensor systems. Firefighters typically follow well-rehearsed search procedures that in detail dictate how the they will move (e.g., left-hand directional searches \cite{Wahlstrom2019b}), and hence, the convergence constraints will often be satisfied within firefighter positioning. In addition, it would be possible to utilize that multiple firefighter teams often search the building using the same search pattern.

Several extensions can be imagined. First, note that the estimation framework could just as well be used to calibrate other parameters than the zero-velocity detection threshold, such as the measurement variance for the ZUPT. Similarly, the estimation framework could also be used to calibrate odometry based on other types of sensors, such as visual odometry. Second, it may possible to implement statistical tests that could answer questions such as ``Does this fragment of odometry comply with the assumptions of the FootSLAM algorithm?'', or ``Have any of the underlying factors, such as gait mode or walking surface, changed to such an extent that the odometry needs to be recalibrated?''. By incorporating such tests into the proposed algorithm, a recalibration could be performed whenever it is possible \emph{and} needed. When testing for changes in underlying factors, it may be useful to divide the data into non-overlapping windows, such that the FootSLAM algorithm converges using data from each separate window. Third, we mention the possibility of merging the proposed algorithm with one of the many proposed methods for adaptive thresholding using gait mode classification. Instead of learning a mapping between IMU-derived classification features and suitable thresholds using ground truth data, we could learn it using FootSLAM. This approach may enable a calibration algorithm that quickly adapts to changes in gait conditions (i.e., an algorithm that does not need to wait for convergence of the FootSLAM algorithm before each new recalibration) but which relies on the output from the FootSLAM algorithm to avoid dependence on ground truth data.

\bibliographystyle{IEEEtran}
\bibliography{refs}

\end{document}